\documentclass[3p,twocolumn]{elsarticle}

\usepackage{graphicx}
\usepackage{color} 
\usepackage{amsmath}
\usepackage{amssymb}
\usepackage{array} 
\usepackage{rotating} 

\usepackage{tikz}
\usetikzlibrary{decorations.markings}

\title{Cluster Dynamics Modeling of Niobium and Titanium Carbide Precipitates}
	
\author[imp,ucas]{Nadezda~Korepanova\corref{cor1}}
\ead{nadezhda\_kv@bk.ru}

\author[rom]{Mihai Dima}

\author[imp,ucas,landa,psi]{Long Gu\corref{cor1}}
\ead{gulong@impcas.ac.cn}

\author[imp,ucas]{Hushan Xu}

\address[imp]{Institute of Modern Physics, Chinese Academy of Sciences, Lanzhou, China}	
\address[ucas]{School of Nuclear Science and Technology, University of Chinese Academy of Sciences, Beijing, China}

\address[rom]{Institute for Physics and Nuclear Engineering, 
Str. Atomistilor 407, P.O. Box MG-6, 
R-76900 Bucharest, Romania}

\address[landa]{School of Nuclear Science and Technology, Lanzhou University, Lanzhou 730000, China}
\address[psi]{Paul Scherrer Institute, Villigen, Switzerland}

\cortext[cor1]{Corresponding authors at: Institute of Modern Physics, Chinese Academy of
	Sciences, Lanzhou, China.
\\
	E-mail addresses: nadezhda\_kv@bk.ru (N. Korepanova), gulong@impcas.ac.cn
(L. Gu).
}

\begin{document}

\begin{abstract}	

	Kinetics of niobium and titanium carbide precipitates in iron has been simulated with cluster dynamics. 
	The simulations, carried out in austenite and ferrite for niobium carbides, respectively in austenite
     for titanium carbide, were analyzed for dependency on 
     temperature, solute concentration and initial cluster 
     distribution. 
	The results are presented for different temperatures and solute concentrations and
     compared to available experimental data.
	They show little impact of initial cluster
     distribution beyond a certain relaxation time and that 
	highly dilute alloys with only monomers present a significantly different behavior than 
	less dilute alloys or alloys with different initial cluster distribution.
\end{abstract}


\maketitle

\section{Introduction}
	
Addition of titanium and niobium to 
steels in metallurgy is conducive 
to titanium/niobium carbide precipitates in solid solution
due to their combination with the carbon present in the steel.
This process limits the
formation of chromium carbides, thereby 
preventing intergranular corrosion \cite{Thorvaldsson1980,Sourmail2002}. 
Additionally, finely dispersed 
carbide precipitates increase alloy strengths both
at low and high temperature~\cite{Sourmail2002,Kong2018}. 
Nuclear-grade steels are required to meet higher, additional
standards, and the question of precipitate dynamics
is raised with respect to them acting as point 
defect recombination centers and sinks for helium, 
which reduces void swelling and helium 
embrittlement \cite{din1.4970SwellCreepModel1,Kesternich1981,Kesternich1985a,Kesternich1986a,Kimoto1986}. 
TiC/NbC precipitates tend
to
stabilize dislocation networks, and hence
enhance creep resistance \cite{din1.4970ThermCreep,Cautaerts2018}. 
Titanium carbides are in particular
attractive in this respect, which led to
the development of 15-15Ti steel in 
the 1970's~\cite{Seran2016} for nuclear reactor applications. 
This steel exhibits excellent
resistance to irradiation swelling and creep and 
has been chosen as structural material for 
several Generation-IV designs.

To simulate the precipitation behavior of carbides
we used cluster dynamics (CD),
which
is an effective method to predict microstructural 
evolution in a material - with little
computing overhead for long time period simulations. 
In this method
polyatomic clusters embedded in the solid solution
exchange solute atoms by absorption or emission. 
Time evolution of cluster distributions 
is computed from differential equations coupled 
through monomer exchanges. 

In the case of iron CD has yielded
good results for Cu precipitates in ferrite \cite{Christien2004} and
MnSiNi precipitates in ferrite-martensitic steel \cite{Ke2018}.

For NbC and TiC precipitates
in steel classical kinetic nucleation theory~(CNT) has
been used \cite{Wang2011,Wang2012,Dutta2001,Perrard2007}, 
TiC precipitate modelling 
focusing more on Precipitation-Temperature-Time (PTT) diagrams, 
rather than on time-evolution of mean radius, 
volume fraction, and number density.

This paper 
is organized as follows:
Section~\ref{sec:Methodology} 
gives a brief description of the CD method;
Section~\ref{sec:Results} presents 
our simulation results for 
niobium carbides in austenite and ferrite and 
for titanium carbide in austenite;
Section~\ref{sec:Results} presents 
a comparison of our results to existing experimental data;
and Section~\ref{sec:Conclusion} summarises our study.

\section{Methodology}
\label{sec:Methodology}

In cluster dynamics alloys are treated as binary systems of
an alloy matrix with clusters of solute atoms.
Clusters grow or shrink through
the absorption, respectively emission of solute atoms. 
Time evolution of solute clusters is
dictated by a set of differential
equations 
(\ref{eq:cd-all}-\ref{eq:cd-mono}) 
which assume only monomers to be mobile. 
Small cluster mobility may result from
common drift of monomers,
an assumption reasonable
in dilute alloys \cite{Mathon1997}.

\begin{align}
	\frac{dC_n}{dt} = \beta_{n-1}C_1C_{n-1}& - (\alpha_n+\beta_nC_1)C_n + \nonumber \\
	&+\, \alpha_{n+1}C_{n+1}\,,
	\quad n \ne 1,
	\label{eq:cd-all}
\end{align}

\begin{align}
	\frac{dC_1}{dt} =  - \, 2\beta_1C_1C_1 &+ \alpha_2C_2 + \nonumber \\
	&+ \sum_{i=2}^{N_{max}} \left[ \alpha_i-\beta_iC_1 \right]C_i\,,
	\label{eq:cd-mono}
\end{align}
where 
	$n$ is     the cluster size,
	$N_{max}$  the maximal cluster size,
	$C_n$      the uniform concentration of size $n$ clusters,
	$C_1$      the concentration of monomers,
	$\alpha_n$ the rate of monomer emission from size $n$ clusters,
	$\beta_n$  the rate of monomer absorption by size $n$ clusters,
	the latter two calculated as:
	
	\begin{align}
	\beta_n = 4\pi r_n D/V_{at}^m\,,
	\label{eq:beta}
	\end{align}
	
	\begin{align}
	\alpha_n = \beta_{n-1} C_{eq}^{sol} \exp \left[ \frac{A \left[ \sigma n^{2/3} - \sigma (n-1)^{2/3} \right]}{kT} \right] \,,
	\end{align}
	where 
	$V_{at}^m$ is the atomic volume of the alloy-matrix,
	$r_n$         the radius of size $n$ clusters,
	$D$           the thermal diffusion coefficient of solute 
                      atoms in the system,
	$A$           the geometrical factor, 
	$\sigma$      the interfacial energy between precipitates and matrix,
	$C_{eq}^{sol}$ the equilibrium concentration of solute atoms in the
                       system,
	$T$            the temperature in degrees Kelvin, and
	$k$            the Boltzmann constant.
	The radius of size $n$ clusters is:
\begin{equation}
r_n = \left( \frac{3nV_{at}^p}{4\pi}\right)^{1/3}\,.
\end{equation}

CD is a computationally efficient method, however with
increasing $N_{max}\gtrsim 100$ it can become CPU wise
prohibitive. A 
traditional way to overcome this
is to transform the differential
equations into a 
Fokker-Planck
partial differential equation \cite{FokkerPlank1,FokkerPlank2}:

\begin{align}
\frac{\partial C_n}{\partial t} = -\frac{\partial}{\partial n}& \left[ (\beta_nC_1-\alpha_n)C_n \right] + \nonumber \\
&+ \frac{1}{2} \frac{\partial^2}{\partial n^2}  \left[ (\beta_nC_1+\alpha_n)C_n \right] \,,
\label{eq:FP}
\end{align}

The discretization of the Fokker-Planck equation using
the central difference method brings
Eq.\ref{eq:FP} into the following form:

\begin{align}
\frac{dC_{n_j}}{dt} &= \frac{C_{n_{j-1}}}{n_{j+1}-n_{j-1}} \times \nonumber \\
& \quad \times \left[ (\beta_{n_{j-1}}C_1 - \alpha_{n_{j-1}}) + \frac{\beta_{n_{j-1}}C_1 + \alpha_{n_{j-1}}}{n_j-n_{j-1}} \right] - \nonumber \\
&-\frac{C_{n_j}}{n_{j+1}-n_{j-1}} (\beta_{n_j}C_1 + \alpha_{n_j}) \times  \nonumber \\
& \quad \times \left[ \frac{1}{n_{j+1}-n_j} + \frac{1}{n_j-n_{j-1}} \right] + \nonumber \\
&+\frac{C_{n_{j+1}}}{n_{j+1}-n_{j-1}} \times  \nonumber \\
& \quad \times \left[ - (\beta_{n_{j+1}}C_1 - \alpha_{n_{j+1}}) + \frac{\beta_{n_{j+1}}C_1 + \alpha_{n_{j+1}}}{n_{j+1}-n_j} \right] \,,
\label{eq:FPapp}
\end{align}
and the evolution of monomer concentration to:
\begin{align}
\frac{dC_1}{dt} = &-2\beta_1C_1C_1 +\alpha_2C_2 + \sum_{j=2}^{N_{tr}} \left[ \alpha_j-\beta_jC_1 \right]C_j \, + \nonumber \\
&+ \sum_{j>N_{tr}}^{N_{max}} \left[ \alpha_j-\beta_jC_1 \right]C_j\frac{n_{j+1}-n_{j-1}}{2}  \,.
\end{align}
where $n_j$ is defined as follows:		
\begin{equation}
\left\{ 
\begin{array}{l}
n_j = j \,, \forall j\le N_{tr},	\\
n_j = N_{tr} + \frac{1-\lambda^{j-N_{tr}}}{1-\lambda} \,, \forall j> N_{tr},	\\
\end{array}
\right.
\end{equation}
The above system is reduced to the initial differential equations
for $n_j~=~j$.
This numerical scheme does not strictly conserve matter 
(as do differential equations (\ref{eq:cd-all}) and (\ref{eq:cd-mono})),
however under carefully defined circumstances
the looses are small and acceptable. 
To solve the above system we used in our study
the ODEINT solver~\cite{odeint}.

In this study we assume 
the diffusion coefficient of titanium/niobium carbide
to be determined by the most resistive element, i.e.
- we used the Ti and Nb diffusion coefficients in the
simulation, correspondingly.

\begin{figure}[h!]
	\includegraphics[scale=0.6]{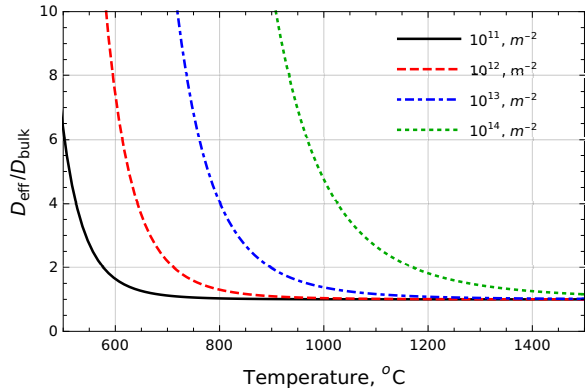}
	\caption{Effect of dislocation on diffusivity - the
ratio of effective diffusivity to bulk diffusivity in austenite
steel for several values of dislocation densities.}
	\label{fig:DeffToDbulkPlot}
\end{figure}

References \cite{Wang2016} and \cite{Dutta2001}
cite the ``pipe-diffusion'' effect for TiC/NbC precipitation kinetics,
respectively a faster
solute diffusion along dislocations than in the lattice in general.
We
included
the effect of dislocations in the model, in Eq.\ref{eq:beta},
as a modified effective diffusivity \cite{Dutta2001}:
\begin{equation}
D_{eff} = D_{disl}\pi R_{core}^2\rho + D_{bulk}(1-\pi R_{core}^2\rho)\,,
\label{eq:Diffusion}
\end{equation}
where 
$D_{disl}$ is the
diffusion along dislocations (equal to $D_{bulk}\alpha_{disl}$, 
with
$\alpha_{disl}$ a correction factor
defined according to \cite{DislocDiff} and 
presented in Table~\ref{tab:DataParameters}), 
$D_{bulk}$ the bulk diffusivity, 
$R_{core}$ the radius of the dislocation core, and
$\rho$ the dislocation density.
Fig.~\ref{fig:DeffToDbulkPlot} illustrates 
the effect of dislocations on
diffusivity and how this changes with temperature. 
The figure shows the ratio of 
effective diffusivity to bulk diffusivity in austenite 
steel for several values of dislocation densities. 
The ratio increases with dislocation density 
increase and drops sharply with increasing temperature.

The results of the simulation are 
the time-evolution of the
mean radius,
the $\bar{r}$ volume fraction, $f_v$ , 
and the
number density of precipitates, $N_{tot}$ - 
calculated with the following equations:

\begin{itemize}
	\item Mean radius
		\begin{equation}
			\bar{r} = \left( \frac{3\bar{n} V^p_{at}}{4\pi}\right)^{1/3}\,, 
		\end{equation}	
		where $V^p_{at}$ is 
the
atomic volume of precipitate, and $\bar{n}$ 
the mean size of precipitate clusters:
		\begin{equation}
			\bar{n} = \frac{ \sum_{j_{cut} }^{N_{max}} n_j C_{n_j} \Delta n_j}{\sum_{j_{cut}}^{N_{max}}C_{n_j} \Delta n_j} \,,
			\label{eq:meanN}
		\end{equation}
		with
 $\Delta n_j =n_j-n_{j-1}$, $j_{cut}$ 
taken such that $r_{j_{cut}}=1\,nm$ for TEM data
(given the resolution limit in
 \cite{Hansen1980, Kesternich1985}, and $r_{j_{cut}}=0.5\,nm$ for SANS data).

	\item Volume fraction 
		\begin{equation}
			f_v = \frac{V_{at}^{p} }{V_{at}^{mat}} \sum_{j_{cut}}^{N_{max}} n_j C_{n_j} \Delta n_j \times 100\%\,.
		\end{equation}

	\item Number density 
		\begin{equation}
			N_{tot}(t) = \frac{1}{V_{at}^{mat}}\sum_{j_{cut}}^{N_{max}} C_{n_j}(t) \Delta n_j \,,
		\end{equation}
		with $V_{at}^{mat}$ the atomic volume of the matrix.
\end{itemize}

In our study we used an
initial cluster distribution described by:
\begin{equation}
	\left\{
		\begin{array}{l}
			C_1=xC_0\,, \\
			C_n=\frac{(100-x)C_0}{n\sum_{n=1}^{M}\Delta n}\,, \quad 2 \le n \le M \\
			C_n=0\,, \qquad \qquad \quad n> M\\
		\end{array}
	\right.
	\label{eq:InDist}
\end{equation}
where $C_0$ is the concentration of the alloying element in steel, 
$x$ the part of the alloying element in monomer form,
$M$ the maximal cluster size assumed to exist in the steel at
 moment $t=0$. 
In the next section, we show
the dependence of CD results on the initial state of the system.

\section{Results and Discussion}
\label{sec:Results}

In this section, we 
present the results of our CD simulations for NbC and TiC 
in ferrite and austenite and compare them with experimental data
from literature. 
The parameters used in the simulations are 
shown
in Tables~\ref{tab:SimulParameters}~and~\ref{tab:DataParameters}. 
Table~\ref{tab:SimulParameters} displays
the material parameters for TiC, NbC and iron matrix
and 
Table~\ref{tab:DataParameters} gives references 
to experimental data and the conditions under which they
was obtained. 

\begin{table*}[h!]
	\centering
	\footnotesize
	\caption{Material parameters for 
                 Titanium and Niobium carbides and for the Iron matrix.}
	\begin{tabular}{llllll}
		\hline
		&Parameter             & Symbol                 & Units   & Value                                             & Reference            \\ \hline
		TiC&                   &                        &         &                                                   &                      \\
		&Lattice parameter     & $a$                    & $nm$    & 0.433                                             & \cite{Jang2012}      \\
		$\gamma Fe$&           &                        &         &                                                   &                      \\
		&Diffusion coefficient & $D^{Ti}$               & $m^2/s$ & $ 0.15\cdot 10^{-4}\exp\left[ -251000/RT \right]$ & \cite{Moll1959}      \\
		&Interfacial energy    & $\sigma$               & $J/m^2$ & 0.2  & \cite{Gustafson2000}      \\
		&solubility limit      & $\log \left[MC\right]$ &         & $2.97 - 6780/T$                                   & \cite{Dumitresc1999}\\
		\hline
		NbC&                   &                        &         &                                                   &                      \\
		&Lattice parameter     & $a$                    & $nm$    & 0.445                                             & \cite{Zou1991}  \\
		$\gamma Fe$&           &                        &         &                                                   &                      \\
		&Diffusion coefficient & $D^{Nb}$               & $m^2/s$ & $ 0.75\cdot 10^{-4}\exp\left[ -264000/RT \right]$  &  \cite{Kurokawa1983}  \\
		&Interfacial energy    & $\sigma$               & $J/m^2$ & $1.0058 - 0.4493\cdot 10^{-3}T(K^o)$           & \cite{Yong1989}      \\
		&solubility limit      & $\log \left[MC\right]$ &         & $2.06 - 6770/T$    &\\
		$\alpha Fe$&           &                        &         &                                                   &                      \\
		&Diffusion coefficient & $D^{Nb}$               & $m^2/s$ & $ 1.27\cdot 10^{-5}\exp\left[ -224000/RT \right]$ & \cite{Perrard2007}      \\
		&Interfacial energy    & $\sigma$               & $J/m^2$ & 0.5              & \cite{Perrard2007}      \\
		&solubility limit      & $\log \left[MC\right]$ &         & $5.43 - 10960/T$                     & \cite{Perrard2007}
		\\ \hline
		Matrix&                   &                        &         &                                                   &                      \\
		&Lattice parameter     & $a_{\gamma Fe}$                    & $nm$    & 0.358                           & \\
		&Lattice parameter     & $a_{\alpha Fe}$                    & $nm$    & 0.287                           & \\
		&Correction factort& $\alpha_{disl}(\gamma Fe)$&& $0.643\exp(118700/(R*T))$& \cite{DislocDiff}\\
		&Correction factor& $\alpha_{disl}(\alpha Fe)$&& $0.0133\exp(115000/(R*T))$& \cite{DislocDiff}\\
		\hline
	\end{tabular}
	\label{tab:SimulParameters}
\end{table*}

\begin{table*}[h!]
	\centering
	\footnotesize
	\caption{Experimental datasets and the concentrations 
                 and temperatures at which they were measured.}
	\begin{tabular}{llllll}
		\hline
		Reference			
		&Ti/Nb, wt\%	&C, wt\%	&Temperatures,  $^oC$	&Matrix	&Dislocation density, $m^{-2}$	\\ \hline
		TiC&	&	&	&		&\\
		\cite{Kesternich1985}
		&0.31	&0.1	&T=750	&$\gamma Fe$	&$6\cdot10^{14}$ (from \cite{Kesternich1985})	\\ 
		\cite{Wang2013,Wang2016}		
		&0.1	&0.05	&T=925	&$\gamma Fe$	&$3.24\cdot10^{13}$ (calc. from \cite{Wang2016})	\\ 
		\cite{Gustafson2000}
		&0.4	&0.07	&T=900	&$\gamma Fe$	&-- (assumed $6\cdot10^{14}$)\\ 
		\cite{Liu1988}
		&0.25	&0.05	&T=900	&$\gamma Fe$	&-- (assumed $6\cdot10^{14}$)\\ 
		\hline
		NbC&	&	&	&	&		\\
		\cite{Hansen1980}	
		&0.031/0.095	&0.1/0.1		&T=900/950		&$\gamma Fe$	&-- (assumed $10^{11}$)\\
		\cite{Perrard2006,Perrard2007}	
		&0.040/0.079	&0.0058/0.01	&T=600/700/800	&$\alpha Fe$	&$2\cdot10^{14}$ (from \cite{Perrard2006,Perrard2007})	\\
		\hline
	\end{tabular}
	\label{tab:DataParameters}
\end{table*}

\subsection{Niobium Carbide}

\begin{figure}[h!]
	\includegraphics[scale=2]{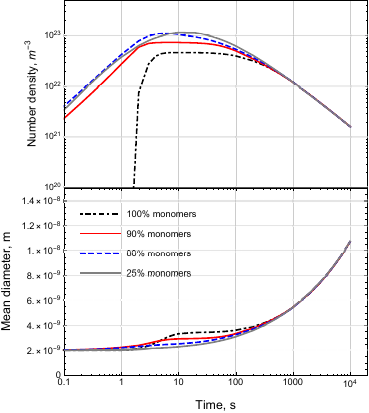}
	\caption{Dependence of simulation results on initial cluster distributions. T=950$^o$C, C(wt\%Nb)=0.095. }
	\label{fig:NbCInitRasp}
\end{figure}

\begin{figure*}[h!]
	\includegraphics[scale=2]{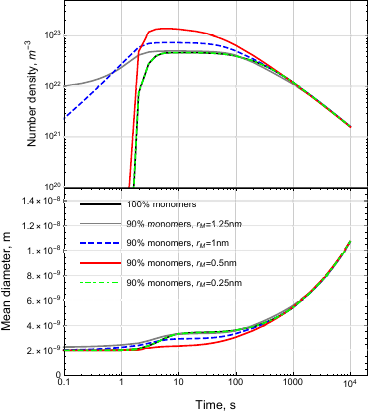} \hspace{0.5cm}
	\includegraphics[scale=2]{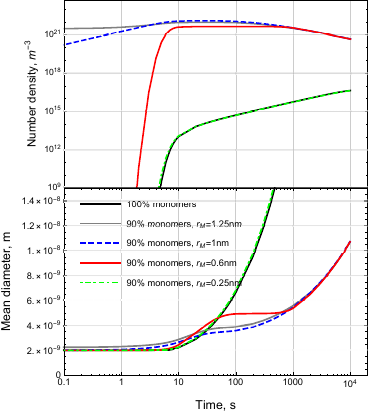}
	\caption{Dependence of simulation results on initial cluster distributions and radius $r_M$. T=950$^o$C, C(wt\%Nb)=0.095 (left graph), C(wt\%Nb)=0.031(right graph). }
	\label{fig:NbCInitRaspRad0031}
\end{figure*}

Figures \ref{fig:NbCInitRasp} 
    and \ref{fig:NbCInitRaspRad0031} 
show the dependence on the initial cluster distribution 
for the
time-evolution of precipitate mean radius and number density. 
We assessed this in order to check the sensitivity of our simulations
to the initial state of the system.
The initial cluster distributions for our simulations
are described by Eq.\ref{eq:InDist}.   
Note in the
figures that we 
index the radius $r_M$ by $M$,
the maximal size of a cluster initially existing in the system.
The cluster distributions we
used were both described by
Eq.\ref{eq:InDist}, as well as
arbitrary distributions: Poisson-like, or step-function.
The simulation results are however the same
for all distributions,
warranting our use of Eq.\ref{eq:InDist} described 
initial distributions throughout our study. 

Figure \ref{fig:NbCInitRasp} shows 
the time-evolution of precipitate 
mean radius and number density in
4 distinct cases
for which we
varied the
concentrations of monomers and other clusters (see Eq.\ref{eq:InDist}),
while keeping $r_M$ constant.
Complementary,
Figure~\ref{fig:NbCInitRaspRad0031} presents
the time-evolution of the mean radius and number density in 
relation to $r_M$, for the same correspoinding monomer concentrations. 
As
comparison 
Figure~\ref{fig:NbCInitRaspRad0031} shows
the simulation only for monomers.

As
evident
in Figures~\ref{fig:NbCInitRasp} -- \ref{fig:NbCInitRaspRad0031}
the 
initial cluster distributions play a
role only in the initial departure time
(with the
notable
exception of the 0.031wt\%Nb-steel simulation,
with monomers and very small clusters). 
After 1000 s the
effect of the initial cluster distribution
washes out
and all simulations become indistinguishable from one another.
The exception
case 
of 0.031wt\%Nb-steel with monomers and very small clusters
is shown in Fig.\ref{fig:NbCInitRaspRad0031}. 
Its
anomalous behaviour was also
observed for TiC in $\gamma$Fe and NbC in $\alpha$Fe.
A possible explanation
could be that
in dilute alloys there are few precipitation centers,
which have the opportunity to grow faster than in 
the case of higher solute concentrations. 
The amount of such clusters remains however
small (see upper Fig.\ref{fig:NbCInitRaspRad0031}). 
Had we introduced higher clusters in the
initial distribution, the precipitate kinetic
would have followed the standard behaviour.

\begin{figure}[h!]
	\centering
	\includegraphics[scale=0.82]{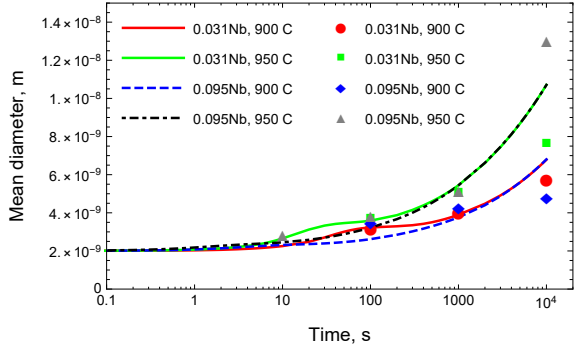}
	\caption{Comparison of simulation 
 results with experimental data for NbC precipitates 
in austenitic stainless steel. 
The dots represent the experimental data of~\cite{Hansen1980}.}
	\label{fig:NbCdataFit}
\end{figure}

Comparing our simulations (Fig.\ref{fig:NbCdataFit})
with experimental data for niobium carbide precipitates
in austenite~\cite{Hansen1980} 
we find that in 0.031wt\%Nb-steel 90\% of niobium should
exist as monomers -  
because if we would accept the 100\% assertion of~\cite{Hansen1980},
the simulations would contradict the experimental data.
We assume\footnote{The exact dislocation density 
for the steel used by Hansen~et~al.~\cite{Hansen1980} is not available.
Hence, we adjusted this parameter to agree
with the experimental data.
The result shows the dislocation density of steel 
to be in the range $10^{11}-10^{12}\,m^{-2}$.
This concurs with the mention that the steel was well 
annealed~\cite{Hansen1980}.
} that the remaining 10\% are distributed in clusters with
$r_M<1$nm is invisible to TEM.

\begin{figure*}[h!]
	\centering
	\includegraphics[scale=2.4]{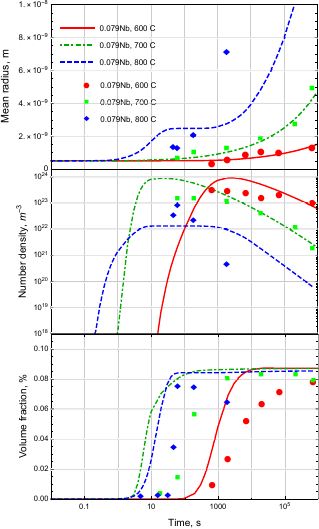}\hspace{0.5cm}
	\includegraphics[scale=2.4]{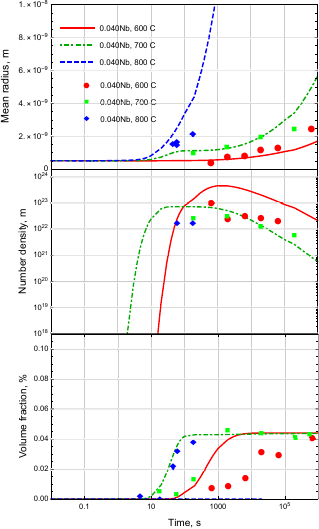}
	\caption{Simulation results for C(wt\%Nb)=0.079~(left) 
and 0.040~(right). 
The dots represent experimental data 
of~\cite{Perrard2006,Perrard2007}. 
For C(wt\%Nb)=0.040 at T=800$^oC$ 
the simulated number density and volume 
fraction of precipitates is too small and, therefore
 invisible in the figures. }
	\label{fig:NbCPerr}
\end{figure*}

For niobium carbide precipitates
in ferrite the simulation results and 
experimental data \cite{Perrard2007,Perrard2006} 
are depicted in Fig.\ref{fig:NbCPerr}. 
Our model with the set of parameters 
from Table~\ref{tab:SimulParameters} matches quite
 well the experimental
data of \cite{Perrard2007,Perrard2006}, less
to some extent 
the
volume fraction, 
predicting faster clustering of the 
precipitates than experimentally observed.
Fig.\ref{fig:NbCPerr}
also shows 
better agreement with 
experimental data at low temperatures versus high temperatures. 
This could be due
to the higher energy available
at high temperature that activates
the diffusion of small clusters along with
monomers, whereas
in our model only the monomers are mobile.

\subsection{Titanium Carbide}

\begin{figure*}[h!]
	\centering
	\includegraphics[scale=0.85]{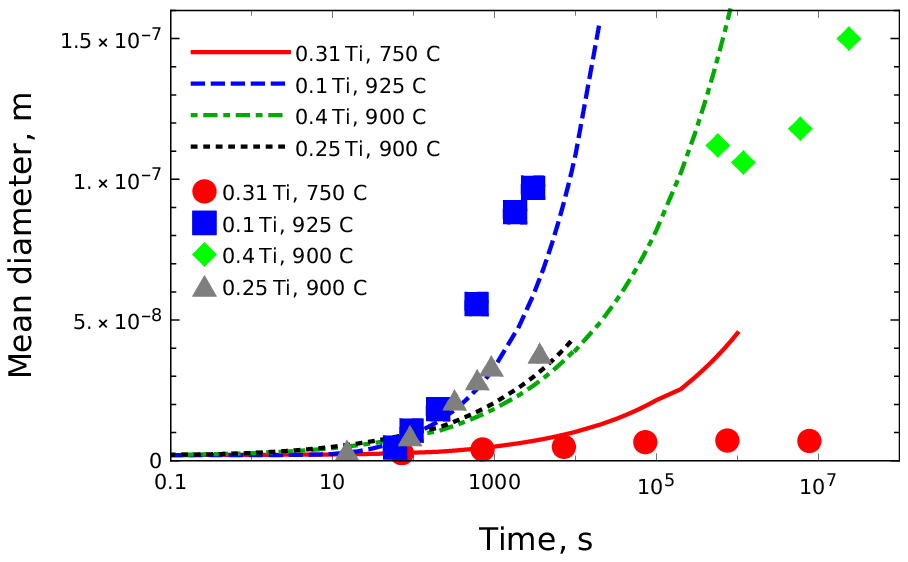} \hspace{0.5cm}
	\includegraphics[scale=0.6]{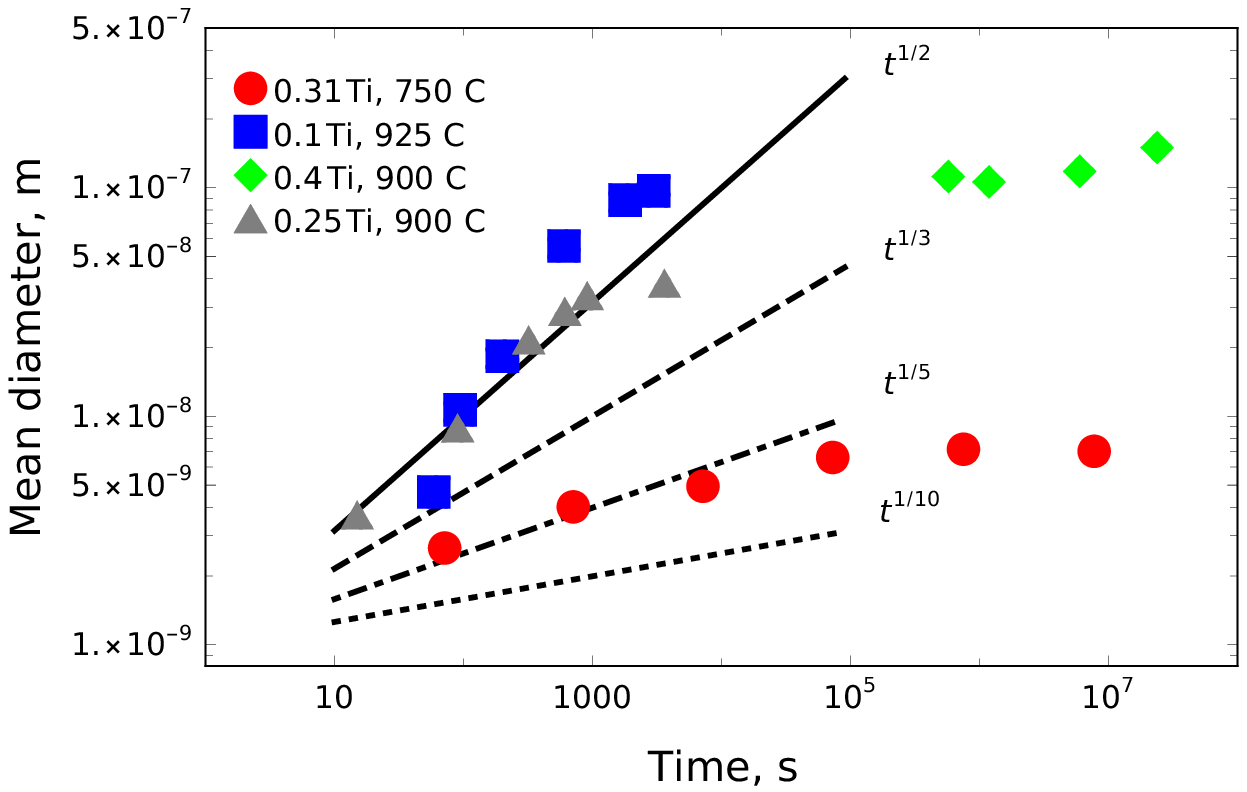} 
	\\  \hspace{1cm} $a)$   \hspace{8cm}   $b)$ 
	\caption{Comparison of simulation results 
with experimental data for TiC precipitates 
in austenitic stainless steel. The dots represent the
experimental data of \cite{Kesternich1985,Gustafson2000,Wang2013,Liu1988}.}
	\label{fig:TiCfitData}
\end{figure*}

\begin{figure*}[h!]
	\centering
	\includegraphics[scale=1.5]{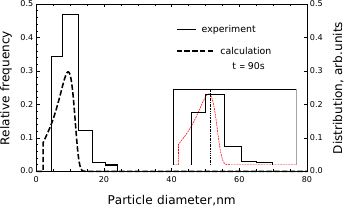}\hspace{0.1cm}
	\includegraphics[scale=1.5]{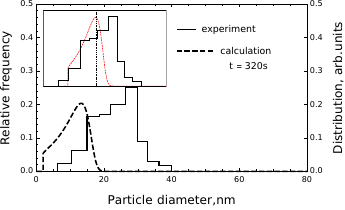}\hspace{0.1cm}
	\includegraphics[scale=1.5]{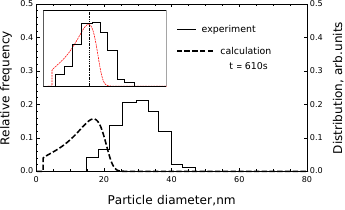}\\
	\vspace{0.3cm}
	\includegraphics[scale=1.5]{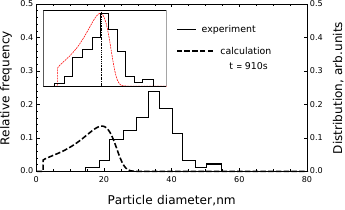}\hspace{0.5cm}
	\includegraphics[scale=1.5]{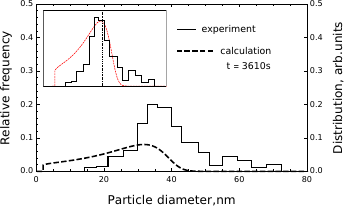}
	\caption{Size distributions from
experiment~\cite{Liu1988} and simulation for 
different times at 900$^oC$. The inserts
inside the graphs display the experimental and 
calculated distributions shifted 
such 
that both have the same mean 
particle diameter. The dot-dashed vertical line represents mean diameter.}
	\label{fig:TiCLiuDistrComparison}
\end{figure*}

\begin{figure*}[h!]
	\centering
	\includegraphics[scale=1.5]{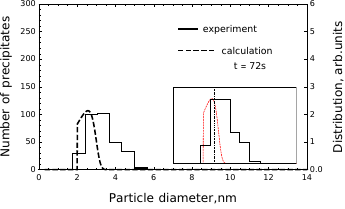}\hspace{0.1cm}
	\includegraphics[scale=1.5]{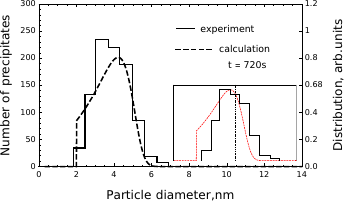}\hspace{0.1cm}
	\includegraphics[scale=1.5]{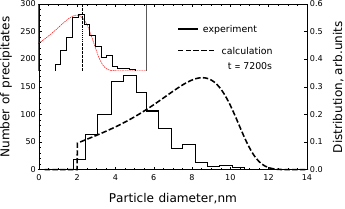}
	\caption{Size distributions from
experiment~\cite{Kesternich1985} and simulation 
for different times at 750$^oC$. The inserts
inside the graphs display the experimental and 
calculated distributions shifted
such that both have the same 
mean particle diameter. The dot-dashed vertical line represents mean diameter.}
	\label{fig:TiCKesternichDistrComparison}
\end{figure*}

Simulation results for TiC precipitates in
austenite and experimental 
data from \cite{Kesternich1985,Gustafson2000,Wang2013,Liu1988}
are presented in 
Figures~\ref{fig:TiCfitData} -- \ref{fig:TiCKesternichDistrComparison}. 
Fig.\ref{fig:TiCfitData} displays the 
time evolution of mean particle diameter - CD
predicting a
particle diameter $\propto t^{1/3}$, regardless
concentration, temperature, and dislocation density 
(except in 0.1wt\%Ti-steel -- $\propto t^{1/2}$). 
Experimental data on the other hand
exhibits two regions:
an initial
 region, with
mean diameter is proportional to time exponent with factor 0.5-1.0;
and a secondary region with
the time exponent having a very low factor, practically
a plateau. 
Although in \cite{Liu1988} the secondary 
region is considered $\propto t^{1/3}$ (reflecting
the Ostwald ripening phenomenon controlled by
 bulk diffusion of Ti), the authors
believe the time exponent has a much lower factor 
(likely due to a different phenomenon). 
The two regions are also clearly observed in the size distributions.

The size distributions from experimental data \cite{Liu1988} 
are shown in  Fig.\ref{fig:TiCLiuDistrComparison}. 
Although our model predicts a smaller mean diameter, 
the overall distribution
shape of the sizes
 is strikingly similar to experimental one
(up to 3610 s, after which
the distributions start to differ).
Additionally,
in the inserts
in Fig. \ref{fig:TiCLiuDistrComparison}
the experimental and simulated size distributions are
plotted such that both have the same mean diameter and maximal magnitude. 
The comparison of size distributions relative
to the experiment~\cite{Kesternich1985} 
at 750$^oC$ (Fig.\ref{fig:TiCKesternichDistrComparison}) 
shows a similar change in size distribution.
The small difference
observed 
might suggest a competing
mechanism controling the growth of TiC precipitates.

In \cite{Kesternich1985} it
was mentioned that the pinning of mobile dislocation 
affects TiC precipitate kinetics in 
the 
temperature range 650\,--\,900$^oC$.
However, CD applies only 
diffusion-controlled growth of precipitate clusters. 
To overcome this obstacle, a model for 
time-evolution of mobile dislocation density should 
be introduced alongside with the dependency of 
diffusion on mobile dislocation density.
The verification of this suggestion is postponed to further study.
Note
that the data itself is very scarce for TiC precipitations. 
We assess
that there is a necessity for more experimental 
and theoretical work on TiC precipitates. 

\section{Conclusion}
\label{sec:Conclusion}

In the present paper
we
applied cluster dynamics to model
 precipitation kinetics of niobium and titanium carbides in steels. 
The kinetic of NbC precipitates have been simulated 
for ferritic and austenitic iron matrices. 
Our simulation
results are in agreement with experimental data. 
We have
analyzed the results for dependency on initial cluster distribution, 
where we considered 
various types of distributions and monomers concentration. 
The analysis has shown that the 
initial distribution plays a role only in the initial-time range. 
After an initialt-time
all simulations follow the same pattern. 
The analysis has also
 shown the ``special'' behavior of precipitates 
if only monomers are present in very dilute alloys: 
the fast growth of mean 
particle diameter, while number density remains small. 
We have suggested that in dilute alloy 
fewer precipitation centers are created 
and they have less competition, 
which allows those centers to grow faster. 

For TiC on the other hand,
 the simulation results and experimental data 
differ more pronounced. This
could be
explained by another controlling mechanism besides diffusion. 
Such mechanism could be mobile dislocation 
and its pining, which was suggested in \cite{Kesternich1985}. 
We
believe, that more experimental and theoretical work 
is needed to correctly model titanium carbide precipitates kinetics.

\section*{Acknowledgments}
N. Korepanova is grateful for the CAS-TWAS President's Fellowship Programme for this doctoral fellowship (2016CTF004).

\bibliographystyle{elsarticle-harv}

\end{document}